\documentclass[aps,twocolumn,preprintnumbers]{revtex4}
\usepackage{setspace}
\usepackage{graphicx}
\usepackage{subfigure}

\preprint{The following article has been submitted to Applied Physics Letters}

\begin{document}

\title{Transit-Time Spin Field-Effect-Transistor}
\author{Ian Appelbaum}
\affiliation{Electrical and Computer Engineering Department,
University of Delaware, Newark, Delaware, 19716}
\author{Douwe Monsma}
\affiliation{Cambridge NanoTech, Inc., Cambridge, MA 02139}

\begin{abstract}
We propose and analyze a four-terminal metal-semiconductor device that uses hot-electron transport through thin ferromagnetic films to inject and detect a charge-coupled spin current transported through the conduction band of an arbitrary semiconductor. This provides the possibility of realizing a spin field-effect-transistor in Si, using electrostatic transit-time control in a perpendicular magnetic field, rather than Rashba effect with spin-orbit interaction.
\end{abstract}

\maketitle
\newpage

When Datta and Das first introduced the spin field-effect-transistor (spin-FET) in 1990, they ignited years of experimental efforts to create a device wherein the electron transport is modulated by electrostatic control of electron spin direction.\cite{C1} They proposed that in the semiconductor channel between a ferromagnetic (FM) source providing spin polarized electrons and a FM spin-analyzing drain, the presence of spin-orbit interaction and an electric field induced by a voltage on a gate will cause a Rashba effect, acting as an effective magnetic field, causing controlled spin precession.\cite{2a,2b} Their analogy to an electro-optic modulator illustrated how the drain current was expected to be periodically dependent on gate voltage. Since this seminal contribution, however, many difficulties in realizing a true spin-FET have arisen. 

First, efficient spin injection was shown to be difficult due to the inherent conductivity mismatch between ferromagnetic metals and typical semiconductors in ohmic contact.\cite{3} This problem has been solved in recent years by using a tunnel-junction barrier between the FM and semiconductor,\cite{4,5,6} or with ballistic hot-electron transport through the FM and injection over a Schottky barrier into the semiconductor conduction band.\cite{7}

Although optical methods have been used to reveal much of the underlying physics of spin dynamics and transport in direct-bandgap semiconductors,\cite{8} electrical spin detection remains a significant problem, especially for indirect-bandgap semiconductors. Recent progress using the epitaxial Fe/GaAs system for non-local detection of spin diffusion with voltage sensing is encouraging,\cite{9} but a device not requiring epitaxial interfaces would clearly be more useful in adapting spin transport to other semiconductors. One semiconductor of interest in which spin transport has not been studied extensively is Si, which has several promising attributes suggesting a long spin lifetime: zero nuclear spin for the most abundant isotope Si$^{28}$, crystal lattice inversion symmetry maintaining spin-degenerate bands, and low spin-orbit interaction.\cite{10} The latter quality, while important for long spin lifetime, is contrary to the initial suggestions of Datta and Das, because it leads to a weak Rashba effect in an applied perpendicular electric field. Therefore, any spin-FET based on Si must depart from the Datta-Das design and make use of a mechanism of spin precession other than Rashba effect. Here we describe a semiconductor device that uses hot-electron transport in FM thin films for spin polarization and injection, electrostatically controlled spin precession in a semiconductor, and hot-electron spin analysis. The device design is not sensitive to metal-semiconductor interfaces and so is applicable to any non-degenerate semiconductor making Schottky barriers with FM metals, not just epitaxial systems.

\begin{figure}
  \includegraphics[width=8cm,height=7cm]{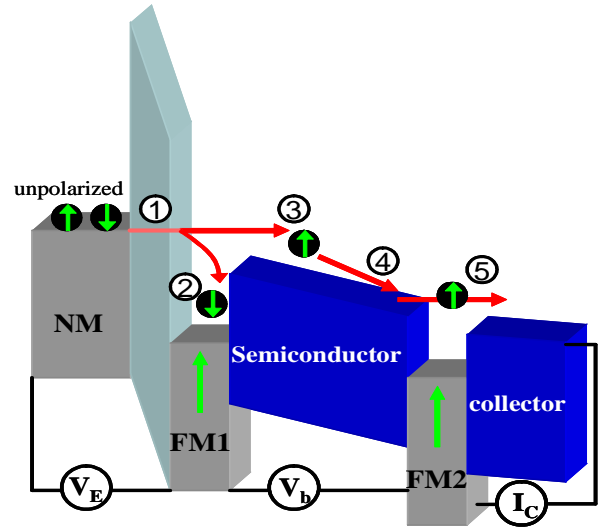}
  \caption{Schematic band diagram of the proposed spin-transport device. Contacts for the emitter bias $V_E$, drift bias $V_b$, and collector current $I_C$ are shown. Refer to text for explanation of transport steps [1]-[5]}
\end{figure}

A schematic band diagram of the proposed device is shown in Figure 1. There are essentially five steps of transport to illustrate. In step 1, a solid-state tunnel junction injects initially unpolarized electrons from the normal metal (NM) emitter through the ferromagnetic base (FM1) by biasing with an emitter voltage ($V_E$). Ballistic spin filtering, as in a spin-valve-transistor,\cite{11,12,13} selectively scatters minority spin hot electrons in FM1 (i.e. those with spin direction antiparallel to the FM1 magnetization direction) (step 2), so that the electrons transported over the Schottky barrier into the conduction band are spin polarized (step 3). After relaxation to the conduction band minimum and drift/diffusion through the semiconductor (step 4), the electrons are ejected over the Schottky barrier between the semiconductor and analyzing ferromagnetic film (FM2). Ballistic spin filtering scatters minority spin electrons, making the collector current ($I_C$) of electrons ballistically transported through FM2, and into the collector conduction band (step 5), dependent on the relative orientation between the electron spin after transport and the FM2 magnetization. This device can be fabricated in a vertical geometry with standard vacuum wafer bonding during FM deposition, as in the original spin-valve transistor.\cite{12}

The semiconductor drift region is bounded on both sides by rectifying Schottky barriers, so the drift bias ($V_b$) does not cause a spurious current to flow. In fact, because this device uses hot-electron transport on both the injection and detection sides, thermalized electrons are excluded. Therefore, the current and voltage in the drift semiconductor are essentially independent during operation. This enables the electric field provided by $V_b$ to directly alter the transit time of spin-polarized electrons from injection at the FM1/semiconductor interface to detection at the FM2/semiconductor interface. If a magnetic field is further provided in the direction of the electron flow (left to right in Fig.1), it will cause precession if it is sufficiently small ($< \approx$1T) that the magnetizations of FM1 and FM2 remain in-plane (vertical in Fig. 1) due to shape anisotropy. Under these conditions of perpendicular magnetic field, the final precession angle will be determined by $\theta=\omega\tau$, where $\omega=g\frac{\mu_B}{\hbar}B$ is the spin precession (Larmor) radial frequency, and $\tau$ is the electron transit time from injection to detection. The fundamental parameters $\hbar$ and $\mu_B$ are the reduced Planck's constant and the Bohr magneton, respectively; $g$ is the spin gyromagnetic ratio and $B$ is the applied perpendicular magnetic field. If transport is dominated by drift, the transit time $\tau$ is $L/v$, where $L$ is the distance from injection to detection and $v$ is the drift velocity $v=\mu E$, where $\mu$ is the electron mobility in the semiconductor and $E$ is the electric field. The applied voltage $V_b$ therefore controls the spin precession angle at the polarization analyzer through its associated electric field $V_b/L$:

\begin{equation}
\theta=g\frac{\mu_B}{h}B\frac{L^2}{\mu V_b}
\end{equation}
             
We can model the collector current dependence on both magnetic and electric fields using 

\begin{equation}
\Delta I_C \propto \int^{\infty}_{0} P(x=L,t)cos(\omega t)e^{-t/\tau_{sf}}dt
\end{equation}

\begin{figure}
  \includegraphics[width=8cm,height=7cm]{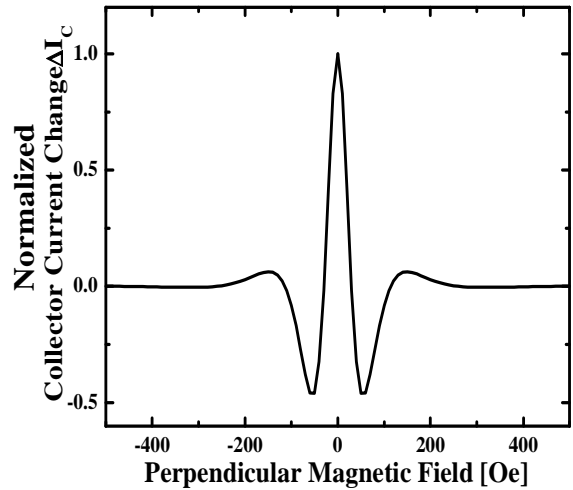}
  \caption{Perpendicular magnetic field dependence of change in collector current ( $\Delta I_C$) obtained by integration of Eq. 2 using Eq. 3 for drift velocity $v\approx 0$, where diffusion is dominant.}
\end{figure}

\begin{figure}
  \includegraphics[width=8cm,height=10cm]{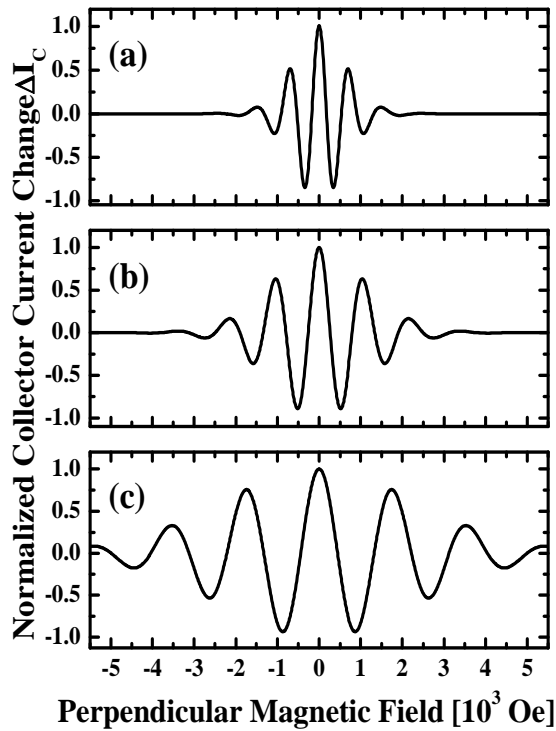}
  \caption{Change in collector current ($I_C$) as a function of magnetic field perpendicular to the FM1 and FM2 magnetization direction.  As the magnetic field increases, the final average precession angle increases in a fixed transit time determined by the drift velocity. (a)-(c) are for drift velocities $v=2\times 10^6$, $3\times 10^6$, and $5\times 10^6$ cm/s, respectively. Since $I_C$ is proportional to the projection of electron spin on the FM2 magnetization direction, the signal oscillates with magnetic field.}
\end{figure}

where $P(x=L,t)$ is the distribution function of electrons arriving at the analyzer ($x=L$) at time $t$, after coherent spin-polarized injection at $t=0$. The exponentially decaying component of the integrand takes into account a finite spin-lifetime, $\tau_{sf}$. This expression has been used to model spin transport in Al using a non-local geometry, but in that case the distribution function is the solution to the standard diffusion equation.\cite{14} In the present case involving diffusion and drift, we must therefore use the solution to the drift-diffusion equation. A convenient approximation to the exact solution of this partial differential equation is a moving Gaussian:

\begin{equation}
P(x,t)=\frac{1}{2\sqrt{\pi D t}}e^{-\frac{(x-vt)^2}{4Dt}}
\end{equation}

where $D$ is the diffusion coefficient. We integrate Eq. 2 using the distribution function given in Eq. 3 as a function of perpendicular magnetic field $B$, when $g=2$, $D=36 cm^2/s$ (parameters for Si), $L=10 \mu m$, and  $\tau_{sf}$ is estimated at 1 ns. For $v\approx 0$ (shown in Fig. 2), the standard diffusion-only case is recovered and the resulting curve is reminiscent of the results shown in Fig. 3 of Ref.[14]. At zero magnetic field, no spin precession occurs, and the spin polarization at the detector is parallel to the FM magnetization if the polarizer and analyzer FM are parallel, causing a high spin detection signal. Since the average transit time over length $L$ for purely diffuse transport is $L^2/2D$, the average electron has precessed $\pm \pi$ radians when the perpendicular magnetic field is on the order of $B_{min}=\pm\frac{Dh}{g\mu_BL^2}$, a condition that causes a pronounced minimum in signal due to the projection of FM2 magnetization on the antiparallel spin direction for small nonzero magnetic field values. Higher-order precession extrema at higher perpendicular magnetic fields are dampened in Fig. 2 by the wide distribution of final spin directions caused by random diffusion in a process called dephasing.

When the drift velocity $v$ is increased by a nonzero $V_b$ to the regime where $L/v\ll L^2/2D$, drift dominates over diffusion. This case is shown in Figure 3(a-c). The shortened transit time $\tau$ causes a reduced precession angle $\theta=\omega\tau$ in a fixed magnetic field, resulting in the $\pm\pi$ precession minima shifted to higher perpendicular magnetic fields. Higher order precession extrema are more apparent as well, due to the reduced role of diffusion and consequently less dephasing.

\begin{figure}
  \includegraphics[width=8cm,height=7cm]{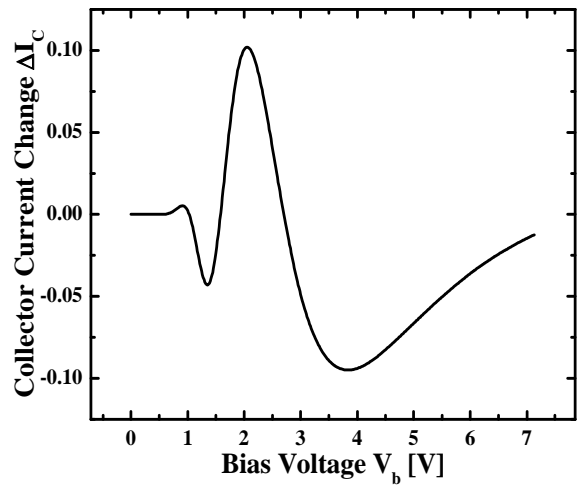}
  \caption{Modulation of collector current $I_C$ with drift voltage bias $V_b$ at a fixed perpendicular magnetic field $B=1000 Oe$. Because the device has rectifying Schottky barriers on both sides of the drift region, this applied voltage only changes the spin-polarized electron transit time without causing spurious currents to flow. As the drift velocity changes, the extrema shown in Fig. 3 pass through this point, causing the oscillations seen here.}
\end{figure}

Now we describe variable electrostatic control over $I_C$ with $V_b$, and demonstrate in principle how to operate this device as a spin FET. If the perpendicular magnetic field is fixed at a nonzero value $\gg B_{min}$, it is apparent from the results shown in Fig. 3 that increasing the accelerating voltage $V_b$ will change the precession angle at detection (x=L) through Eq. 1, and (in minimal dephasing) cause an oscillation in $I_C$. Figure 4 shows the integral Eq. 2 evaluated at a static perpendicular magnetic field $B=1000 Oe$ (which can be supplied in practice by a permanent magnet), as a function of $V_b$, using again $L=10 \mu m$ and  $\mu=1400 cm^2/V\cdot s$. At accelerating voltages $V_b$ close to zero, diffusion is dominant and hence the collector current remains unchanged until $V_b \approx$1V. However, at higher values of voltage bias, the collector current shows several oscillations of increasing magnitude. Comparing with Fig. 3(a-c), it is clear that these oscillations are due to the precession extrema passing through the point where  $B=1000 Oe$ as $v=\mu V_b / L$ changes. This oscillatory dependence on applied accelerating voltage clearly shows that this device is capable of modulating a current with an electrostatic field via controlled precession. It therefore comprises a spin-FET.   

In conclusion, we have proposed a new spin-transport device which can be operated as a spin-FET. The most salient features of this device are (1) ballistic hot-electron transport through FM thin films for both spin polarization and analysis, and (2) the independent control over current and electric field in the drift region provided by Schottky contact at the metal-semiconductor interfaces. These properties allow electrostatic control over the collector current of our device in a static perpendicular magnetic field, and provide a pathway toward investigating spin transport in semiconductors (including Si) previously inaccessible to other spintronics methods.

\bibliography{SMStheory_ver1b}

\begin{thebibliography}{15}
\expandafter\ifx\csname natexlab\endcsname\relax\def\natexlab#1{#1}\fi
\expandafter\ifx\csname bibnamefont\endcsname\relax
  \def\bibnamefont#1{#1}\fi
\expandafter\ifx\csname bibfnamefont\endcsname\relax
  \def\bibfnamefont#1{#1}\fi
\expandafter\ifx\csname citenamefont\endcsname\relax
  \def\citenamefont#1{#1}\fi
\expandafter\ifx\csname url\endcsname\relax
  \def\url#1{\texttt{#1}}\fi
\expandafter\ifx\csname urlprefix\endcsname\relax\def\urlprefix{URL }\fi
\providecommand{\bibinfo}[2]{#2}
\providecommand{\eprint}[2][]{\url{#2}}

\bibitem[{\citenamefont{Datta and Das}(1990)}]{C1}
\bibinfo{author}{\bibfnamefont{S.}~\bibnamefont{Datta}} \bibnamefont{and}
  \bibinfo{author}{\bibfnamefont{B.}~\bibnamefont{Das}},
  \bibinfo{journal}{Appl. Phys. Lett.} \textbf{\bibinfo{volume}{56}},
  \bibinfo{pages}{665} (\bibinfo{year}{1990}).

\bibitem[{\citenamefont{Rashba}(1960)}]{2a}
\bibinfo{author}{\bibfnamefont{E.}~\bibnamefont{Rashba}},
  \bibinfo{journal}{Sov. Phys. Solid State} \textbf{\bibinfo{volume}{2}},
  \bibinfo{pages}{1109} (\bibinfo{year}{1960}).

\bibitem[{\citenamefont{Bychov and Rashba}(1984)}]{2b}
\bibinfo{author}{\bibfnamefont{Y.~A.} \bibnamefont{Bychov}} \bibnamefont{and}
  \bibinfo{author}{\bibfnamefont{E.}~\bibnamefont{Rashba}},
  \bibinfo{journal}{J. Phys. C} \textbf{\bibinfo{volume}{17}},
  \bibinfo{pages}{6093} (\bibinfo{year}{1984}).

\bibitem[{\citenamefont{Schmidt et~al.}(2000)\citenamefont{Schmidt, Ferrand,
  Molenkamp, Filip, and van Wees}}]{3}
\bibinfo{author}{\bibfnamefont{G.}~\bibnamefont{Schmidt}},
  \bibinfo{author}{\bibfnamefont{D.}~\bibnamefont{Ferrand}},
  \bibinfo{author}{\bibfnamefont{L.}~\bibnamefont{Molenkamp}},
  \bibinfo{author}{\bibfnamefont{A.}~\bibnamefont{Filip}}, \bibnamefont{and}
  \bibinfo{author}{\bibfnamefont{B.}~\bibnamefont{van Wees}},
  \bibinfo{journal}{Phys. Rev. B} \textbf{\bibinfo{volume}{62}},
  \bibinfo{pages}{R4790} (\bibinfo{year}{2000}).

\bibitem[{\citenamefont{Jiang et~al.}(2005)\citenamefont{Jiang, Wang, Shelby,
  Macfarlane, Bank, Harris, and Parkin}}]{4}
\bibinfo{author}{\bibfnamefont{X.}~\bibnamefont{Jiang}},
  \bibinfo{author}{\bibfnamefont{R.}~\bibnamefont{Wang}},
  \bibinfo{author}{\bibfnamefont{R.~M.} \bibnamefont{Shelby}},
  \bibinfo{author}{\bibfnamefont{R.~M.} \bibnamefont{Macfarlane}},
  \bibinfo{author}{\bibfnamefont{S.~R.} \bibnamefont{Bank}},
  \bibinfo{author}{\bibfnamefont{J.~S.} \bibnamefont{Harris}},
  \bibnamefont{and} \bibinfo{author}{\bibfnamefont{S.~S.~P.}
  \bibnamefont{Parkin}}, \bibinfo{journal}{Phys. Rev. Lett.}
  \textbf{\bibinfo{volume}{94}}, \bibinfo{pages}{056601}
  (\bibinfo{year}{2005}).

\bibitem[{\citenamefont{Smith and Silver}(2002)}]{5}
\bibinfo{author}{\bibfnamefont{D.~L.} \bibnamefont{Smith}} \bibnamefont{and}
  \bibinfo{author}{\bibfnamefont{R.}~\bibnamefont{Silver}},
  \bibinfo{journal}{Phys. Rev. B} \textbf{\bibinfo{volume}{64}},
  \bibinfo{pages}{045323} (\bibinfo{year}{2002}).

\bibitem[{\citenamefont{Hanbicki et~al.}(2002)\citenamefont{Hanbicki, Jonker,
  Itskos, Kioseoglou, and Petrou}}]{6}
\bibinfo{author}{\bibfnamefont{A.}~\bibnamefont{Hanbicki}},
  \bibinfo{author}{\bibfnamefont{B.}~\bibnamefont{Jonker}},
  \bibinfo{author}{\bibfnamefont{G.}~\bibnamefont{Itskos}},
  \bibinfo{author}{\bibfnamefont{G.}~\bibnamefont{Kioseoglou}},
  \bibnamefont{and} \bibinfo{author}{\bibfnamefont{A.}~\bibnamefont{Petrou}},
  \bibinfo{journal}{Appl. Phys. Lett.} \textbf{\bibinfo{volume}{80}},
  \bibinfo{pages}{1240} (\bibinfo{year}{2002}).

\bibitem[{\citenamefont{Jiang et~al.}(2003)\citenamefont{Jiang, Wang, van
  Dijken, Shelby, Macfarlane, Solomon, Harris, and Parkin}}]{7}
\bibinfo{author}{\bibfnamefont{X.}~\bibnamefont{Jiang}},
  \bibinfo{author}{\bibfnamefont{R.}~\bibnamefont{Wang}},
  \bibinfo{author}{\bibfnamefont{S.}~\bibnamefont{van Dijken}},
  \bibinfo{author}{\bibfnamefont{R.}~\bibnamefont{Shelby}},
  \bibinfo{author}{\bibfnamefont{R.}~\bibnamefont{Macfarlane}},
  \bibinfo{author}{\bibfnamefont{G.~S.} \bibnamefont{Solomon}},
  \bibinfo{author}{\bibfnamefont{J.}~\bibnamefont{Harris}}, \bibnamefont{and}
  \bibinfo{author}{\bibfnamefont{S.~S.~P.} \bibnamefont{Parkin}},
  \bibinfo{journal}{Phys. Rev. Lett.} \textbf{\bibinfo{volume}{90}},
  \bibinfo{pages}{256603} (\bibinfo{year}{2003}).

\bibitem[{\citenamefont{Kikkawa and Awschalom}(1998)}]{8}
\bibinfo{author}{\bibfnamefont{J.}~\bibnamefont{Kikkawa}} \bibnamefont{and}
  \bibinfo{author}{\bibfnamefont{D.}~\bibnamefont{Awschalom}},
  \bibinfo{journal}{Phys. Rev. Lett.} \textbf{\bibinfo{volume}{80}},
  \bibinfo{pages}{4313} (\bibinfo{year}{1998}).

\bibitem[{\citenamefont{Lou et~al.}(2007)\citenamefont{Lou, Adelmann, Crooker,
  Garlid, Zhang, Reddy, Flexner, Palmstrom, and Crowell}}]{9}
\bibinfo{author}{\bibfnamefont{X.}~\bibnamefont{Lou}},
  \bibinfo{author}{\bibfnamefont{C.}~\bibnamefont{Adelmann}},
  \bibinfo{author}{\bibfnamefont{S.}~\bibnamefont{Crooker}},
  \bibinfo{author}{\bibfnamefont{E.~S.} \bibnamefont{Garlid}},
  \bibinfo{author}{\bibfnamefont{J.}~\bibnamefont{Zhang}},
  \bibinfo{author}{\bibfnamefont{S.}~\bibnamefont{Reddy}},
  \bibinfo{author}{\bibfnamefont{S.}~\bibnamefont{Flexner}},
  \bibinfo{author}{\bibfnamefont{C.}~\bibnamefont{Palmstrom}},
  \bibnamefont{and} \bibinfo{author}{\bibfnamefont{P.}~\bibnamefont{Crowell}},
  \bibinfo{journal}{Nature Physics} \textbf{\bibinfo{volume}{3}},
  \bibinfo{pages}{197} (\bibinfo{year}{2007}).

\bibitem[{\citenamefont{Zutic et~al.}(2006)\citenamefont{Zutic, Fabian, and
  Erwin}}]{10}
\bibinfo{author}{\bibfnamefont{I.}~\bibnamefont{Zutic}},
  \bibinfo{author}{\bibfnamefont{J.}~\bibnamefont{Fabian}}, \bibnamefont{and}
  \bibinfo{author}{\bibfnamefont{S.}~\bibnamefont{Erwin}},
  \bibinfo{journal}{Phys. Rev. Lett.} \textbf{\bibinfo{volume}{97}},
  \bibinfo{pages}{026602} (\bibinfo{year}{2006}).

\bibitem[{\citenamefont{Monsma et~al.}(1995)\citenamefont{Monsma, Lodder,
  Popma, and Dieny}}]{11}
\bibinfo{author}{\bibfnamefont{D.}~\bibnamefont{Monsma}},
  \bibinfo{author}{\bibfnamefont{J.}~\bibnamefont{Lodder}},
  \bibinfo{author}{\bibfnamefont{T.}~\bibnamefont{Popma}}, \bibnamefont{and}
  \bibinfo{author}{\bibfnamefont{B.}~\bibnamefont{Dieny}},
  \bibinfo{journal}{Phys. Rev. Lett.} \textbf{\bibinfo{volume}{74}},
  \bibinfo{pages}{5260} (\bibinfo{year}{1995}).

\bibitem[{\citenamefont{Monsma et~al.}(1998)\citenamefont{Monsma, Vlutters, and
  Lodder}}]{12}
\bibinfo{author}{\bibfnamefont{D.~J.} \bibnamefont{Monsma}},
  \bibinfo{author}{\bibfnamefont{R.}~\bibnamefont{Vlutters}}, \bibnamefont{and}
  \bibinfo{author}{\bibfnamefont{J.}~\bibnamefont{Lodder}},
  \bibinfo{journal}{Science} \textbf{\bibinfo{volume}{281}},
  \bibinfo{pages}{407} (\bibinfo{year}{1998}).

\bibitem[{\citenamefont{Mizushima et~al.}(1997)\citenamefont{Mizushima, Kinno,
  Yamauchi, and Tanak}}]{13}
\bibinfo{author}{\bibfnamefont{K.}~\bibnamefont{Mizushima}},
  \bibinfo{author}{\bibfnamefont{T.}~\bibnamefont{Kinno}},
  \bibinfo{author}{\bibfnamefont{T.}~\bibnamefont{Yamauchi}}, \bibnamefont{and}
  \bibinfo{author}{\bibfnamefont{K.}~\bibnamefont{Tanak}},
  \bibinfo{journal}{IEEE Trans. Magn.} \textbf{\bibinfo{volume}{33}},
  \bibinfo{pages}{3500} (\bibinfo{year}{1997}).

\bibitem[{\citenamefont{Jedema et~al.}(2002)\citenamefont{Jedema, Heersche,
  Filip, Baselmans, and van Wees}}]{14}
\bibinfo{author}{\bibfnamefont{F.}~\bibnamefont{Jedema}},
  \bibinfo{author}{\bibfnamefont{H.}~\bibnamefont{Heersche}},
  \bibinfo{author}{\bibfnamefont{A.}~\bibnamefont{Filip}},
  \bibinfo{author}{\bibfnamefont{J.}~\bibnamefont{Baselmans}},
  \bibnamefont{and} \bibinfo{author}{\bibfnamefont{B.}~\bibnamefont{van Wees}},
  \bibinfo{journal}{Nature} \textbf{\bibinfo{volume}{416}},
  \bibinfo{pages}{713} (\bibinfo{year}{2002}).

\end{thebibliography}

\end{document}